\documentclass[prl,twocolumn,english,superscriptaddress,floatfix,longbibliography]{revtex4-2}
\usepackage{graphicx}
\usepackage{xcolor}
\newcommand\hidden[1]{} 
\usepackage{mathtools}
\usepackage{babel,blindtext}
\newtagform{hidden}[\hidden]{}{}
\usepackage{amsmath,amssymb}
\usepackage{soul}
\usepackage{float}

\newcommand{\ket}[1]{|#1\rangle}

\begin{document}
\setlength{\pdfpagewidth}{8.5in}
\setlength{\pdfpageheight}{11in}
\title{Observing Parity Time Symmetry Breaking in a Josephson Parametric Amplifier
}
\author{Chandrashekhar Gaikwad}\email{chandrashekhar@wustl.edu}
\affiliation{Department of Physics, Washington University, St. Louis, Missouri 63130, USA}
\author{Daria Kowsari}
\affiliation{Department of Physics, Washington University, St. Louis, Missouri 63130, USA}
\affiliation{Department of Physics \& Astronomy, Dornsife College of Letters, Arts, \& Sciences, University of Southern California, Los Angeles, CA 90089, USA }
\affiliation{Center for Quantum Information Science \& Technology, University of Southern California, Los Angeles, CA 90089, USA}
\author{Weijian Chen}
\affiliation{Department of Physics, Washington University, St. Louis, Missouri 63130, USA}
\author{Kater W. Murch}\email{murch@physics.wustl.edu}
\affiliation{Department of Physics, Washington University, St. Louis, Missouri 63130, USA}

\date{\today}
\begin{abstract}
A coupled two-mode system with balanced gain and loss is a paradigmatic example of an open quantum system that can exhibit real spectra despite being described by a non-Hermitian Hamiltonian. We utilize a degenerate parametric amplifier operating in three-wave mixing mode to realize such a system of balanced gain and loss between the two quadrature modes of the amplifier. By examining the time-domain response of the amplifier, we observe a characteristic transition from real-to-imaginary energy eigenvalues associated with the Parity-Time-symmetry-breaking transition.
\end{abstract}
\maketitle

\begin{figure}
\centering
\includegraphics[width=0.5\textwidth]{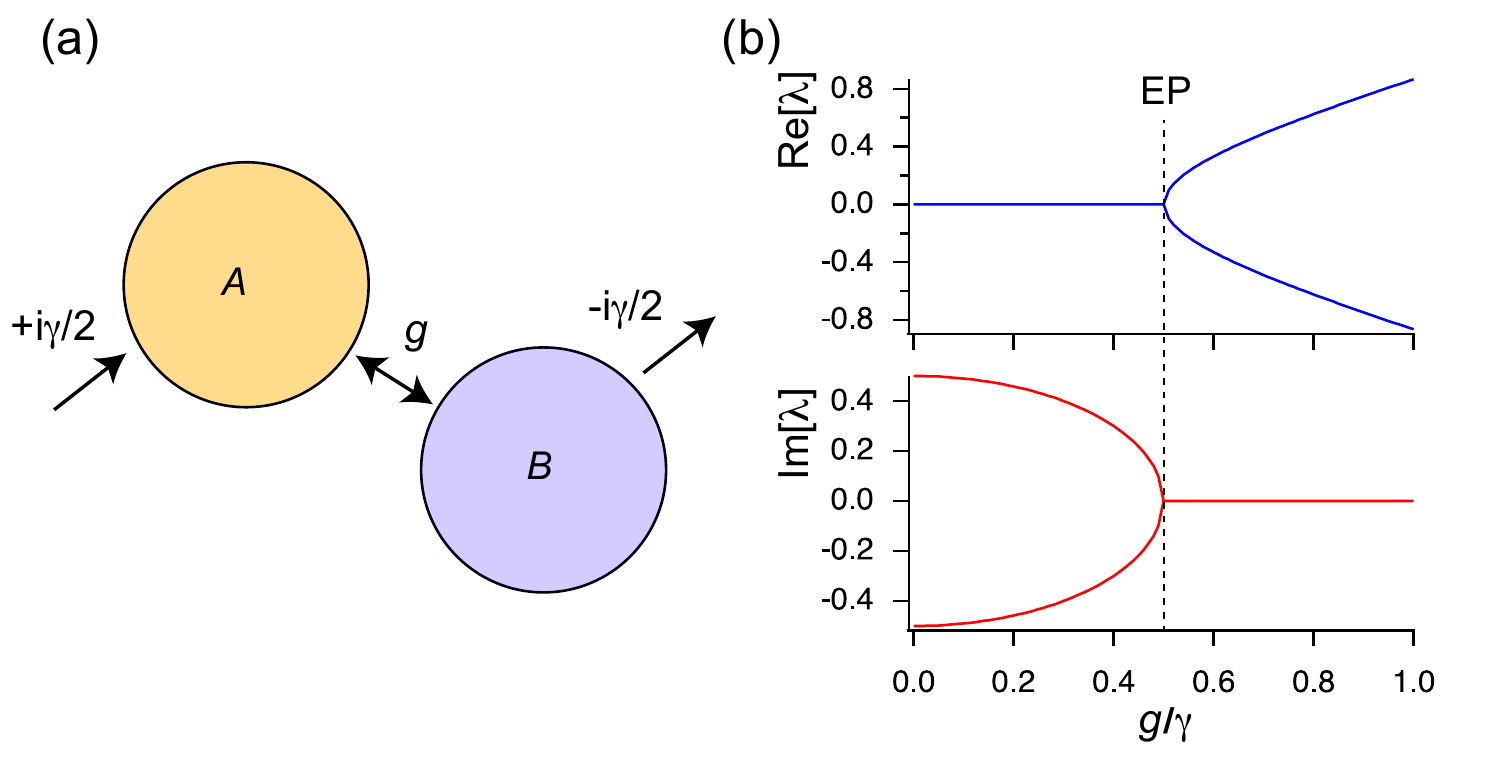}
\caption{{\bf The PT-dimer.} (a) The PT-dimer consists of two coupled modes with respective gain and loss. (b) The complex eigenvalue spectrum of the PT-dimer exhibits a transition from purely imaginary to purely real eigenvalues separated by an exceptional point. For the plot, $\gamma=1$. }
\label{fig:ptdimer}
\end{figure}

Parity-Time (PT) symmetry was introduced as a compelling paradigm for open quantum systems described by non-Hermitian Hamiltonians that can still have real energy spectra \cite{Bend98}.  Originally introduced in the context of complex potentials, the study of PT-symmetry has taken on important relevance due to progress in optics \cite{el2018}.  Figure~\ref{fig:ptdimer} displays a paradigmatic example of the PT-dimer consisting of two modes with balanced gain and loss. 
Originating in optics, such coupled mode systems have been studied extensively in a range of experimental platforms \cite{guo09,rute10,peng14pt,hoda14, zeun15,Li2019,xiao17}. Of particular interest are exceptional point degeneracies \cite{zdem19,Miri19}, which have been identified to confer interesting advantages and functionalities, such as enhanced sensitivity \cite{wier14,hoda17,chen17} and tuning capabilities \cite{Xu16}. Recently, there have been efforts to extend the study of exceptional points to the quantum domain with 
 recent superconducting circuit experiments utilizing purely lossy dynamics to realize passive PT-symmetry \cite{nagh19,Chen21,Chen22,Abb22}. Alternatively, Hamiltonian dilation \cite{Wang2020} can be used to simulate arbitrary Hamiltonians, a method that has been employed with nitrogen-vacancy centers \cite{Wu2019, Liu21}.  
In the quantum domain, amplification is constrained by the requirement that the commutation relation between operators is maintained---leading to fundamental limits for added quantum noise in amplifiers \cite{Clerk2010}. In particular, dissipation-free gain, and therefore noiseless amplification can be achieved in the context of squeezing---where one quadrature is amplified and its conjugate is deamplified, preserving the operator commutation relation. In this Letter, we investigate the realization of a PT-dimer system in this dissipation-free setting \cite{Yue2019,Wang2019,Luo2022}. This goes beyond prior work where incoherent gain is associated with added noise. We utilize a parametric amplifier \cite{Aumentado2020,Yamamoto2008,CastellanosBeltran2008} operating in three-wave mixing mode, where 
the two quadratures of an electromagnetic mode are either amplified or squeezed, corresponding to the respective gain and loss in the PT-dimer model.  A detuning between the pump frequency and the amplifier's resonance introduces coupling between the two quadratures, allowing us to observe the interplay between coupling and gain/loss that characterizes the PT-symmetry-breaking transition in the transient response of the amplifier.  This observation opens the door to a new class of quantum microwave devices that harness non-Hermiticity and exceptional points for a variety of purposes including non-reciprocity \cite{Clerk2022}, enhanced sensing \cite{Lau2018,Budich2020}, and the study of novel topological quantum materials \cite{Shindou2013,Peano2016, Peano2016_twpa, chen18_materials,Gnei2022,Nasari2023}.

The PT-dimer consists of a coupled two-mode system with each mode respectively subject to gain and loss. As displayed in Fig.~\ref{fig:ptdimer}(a), we label these two modes, $A$ and $B$, with equal parts respective gain and loss given by rate $\gamma/2$. The system is invariant under the PT operation as the parity operator switches the modes $A$ and $B$ and time reversal exchanges gain and loss. The two modes are coupled at a rate $g$. The time evolution of the PT dimer is given by the equation of motion, 
\begin{equation}
    i\partial_{t} \begin{pmatrix} A\\B \end{pmatrix} = H_\mathrm{PT} \begin{pmatrix} A\\B \end{pmatrix} , 
\end{equation}
where,
\begin{equation}
    H_\mathrm{PT} = \begin{pmatrix} +i\frac{\gamma}{2}&g\\g& -i\frac{\gamma}{2}\end{pmatrix}
    \label{eq:ptdimer}\\ = g \sigma_{x}+i\frac{\gamma}{2}\sigma_{z}.
\end{equation}\\
Examining the eigenvalues of $H_\mathrm{PT}$, 
\begin{equation}
    \lambda_{\pm} = \pm \sqrt{g^{2}- (\gamma /2)^{2}},
\end{equation}
we can see that the coupling rate $g$ can tune the system from a region of ``unbroken'' PT-symmetry ($g>\gamma/2$), where the  eigenvalues are strictly real, to a region of ``broken'' PT-symmetry ($g<\gamma/2$), where the eigenvalues are purely imaginary. The exceptional point occurs at $g = \gamma/2$, where $H_\mathrm{PT}$ is not diagonalizable and there is a single eigenmode of the system [Fig.~\ref{fig:ptdimer}(b)].

In this work, we utilize a Josephson parametric amplifier to realize the essential physics of the PT-dimer. Josephson parametric amplifiers utilize nonlinearity imparted by Josephson junctions in a superconducting circuit to perform wave mixing between a strong classical pump and weak quantum signals \cite{Aumentado2020}. These amplifiers are utilized broadly in the circuit quantum electrodynamics architecture \cite{Blais2021,Kjaergaard2020} for achieving quantum noise limited amplification and have also been employed for research into quantum nonlinear dynamics \cite{Murch2010,Chen2023}. Here we focus on the case of three-wave mixing, where one pump photon at frequency $\omega_\mathrm{p}$ converts into one signal photon at frequency $\omega_\mathrm{s}$ and one idler photon at frequency $\omega_\mathrm{i}$  operating in degenerate mode, where the  signal is degenerate with the idler ($\omega_\mathrm{s} = \omega_\mathrm{i} = \omega_\mathrm{p}/2$). The Hamiltonian of such a parametric amplifier can be expressed in the rotating frame as \cite{Planat2019},
\begin{equation}
     H _\mathrm{DPA} = \delta a^{\dagger}a+ \frac{\nu}{2}(i a^{\dagger 2}-i a^{2}), 
 \end{equation}
where $\delta \equiv (\omega_\mathrm{p}/2-\omega_{0})$ is the detuning from the resonance of the amplifier at frequency $\omega_0$, $\nu$ is the pump strength, and $a\ (a^\dagger)$ is the photon annihilation (creation) operator for photons at the signal frequency. The last two terms in the Hamiltonian imply the three-wave mixing process between two signal photons and one pump photon; since the pump drive is far stronger than any other signal involved in the system, it is treated as classical.

 
 Even though the system consists of a single bosonic mode, $a$ and $a^{\dagger}$ are still coupled through the parametric drive and can be written in vector basis $\ket{a} = (a,a^{\dagger})^{T}$, where $(\cdot)^T$ represents the transpose operation.  The Heisenberg equation of motion can be formally written as,
\begin{equation}
    i \partial _{t} \ket{a} = \begin{pmatrix} \delta & i\nu \\ i \nu & -\delta \end{pmatrix}\ket{a}.
\end{equation}
To illustrate the connection between the degenerate parametric amplifier and the PT-dimer,  we now transform to the quadrature basis $I =( a + a^\dagger)/\sqrt{2}$, $Q = (a - a^\dagger)/i\sqrt{2}$. 
In this basis, the Heisenberg equation of motion is given by,
\begin{equation}
    i \partial _{t} \begin{pmatrix} I\\i Q\end{pmatrix} = \begin{pmatrix} i \nu & \delta \\ \delta & -i \nu \end{pmatrix} \begin{pmatrix} I\\iQ \end{pmatrix}. \label{eq:iqeom}
\end{equation}
This evolution matrix realizes the PT-dimer Hamiltonian  (\ref{eq:ptdimer}). Here, the pump is the source of coherent gain and loss and the detuning is equivalent to the  coupling. This system exhibits a PT-transition when the threshold ($\delta = \nu$) is crossed. At $|\delta|>|\nu|$ the pump drive is too weak to pin the signal's phase hence the system exhibits oscillatory behavior in the two quadratures---corresponding to the PT-symmetry unbroken regime. However, in the case of $|\delta|<|\nu|$, the pump is sufficient to give one quadrature gain while squeezing the other quadrature---corresponding to the PT-symmetry broken regime. 

\begin{figure}
\centering
\includegraphics[width=0.5\textwidth]{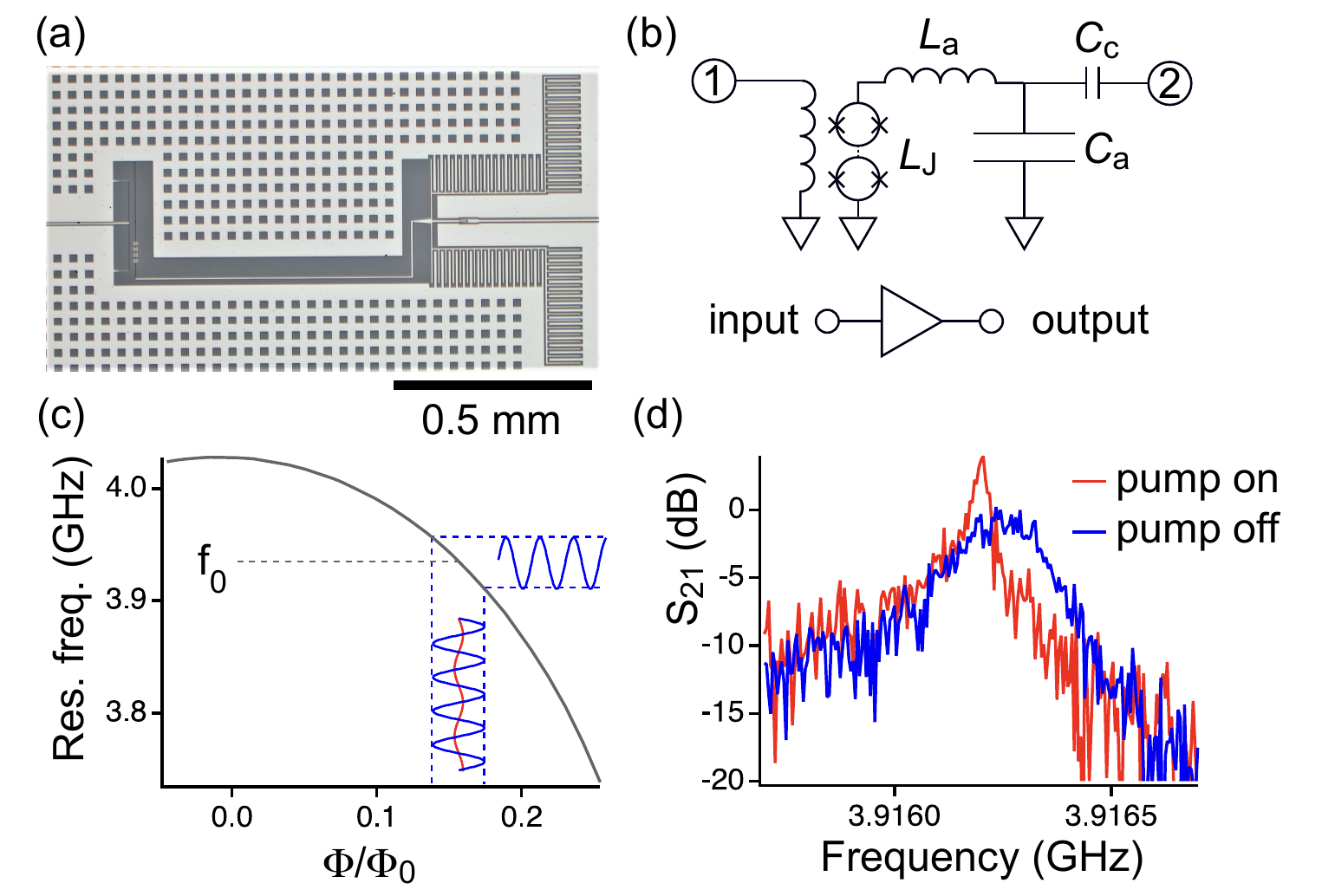}
\caption{{\bf Degenerate parametric amplifier.} (a) Optical micrograph of the device. (b) Circuit schematic of the device; the input port (1) is used for both pump and signal inputs, and the output port (2) is used to monitor the amplifier dynamics. (c) The amplifier is flux biased to generate a first-order flux sensitivity to the pump. When pumped at $\omega_\mathrm{p}$, the amplifier frequency is modulated at $\omega_\mathrm{p}$. We probe the response of a small signal at frequency $\omega_\mathrm{s} = \omega_\mathrm{p}/2$. (d) The amplifier's frequency response exhibits enhanced transmission near the resonance frequency, with amplification above the response when the pump is turned on. }
\label{fig:amp}
\end{figure}

So far, our discussion has focused on Hamiltonian dynamics, yet to probe the associated physics, an experimental device requires input and output ports. These will allow us to inject probe signals, and monitor the dynamics of the quadratures via a weakly coupled output port. Formally, the operators for the output mode are given $a_\mathrm{out} = \sqrt{\kappa_\mathrm{out}} a$, where $\kappa_\mathrm{out}$ is coupling rate to the output port. Hence, the output quadratures $(I_\mathrm{out},Q_\mathrm{out})$ will simply be proportional to the quadratures $(I,Q)$ of the amplifier. Additionally, due to this dissipation of the output port, the associated dynamics will occur in the transient response of the amplifier as it evolves to a steady state. 

We designed a narrow-bandwidth three-wave mixing amplifier.  Figure \ref{fig:amp}a displays the photograph of the device which is patterned as a single-layer device with Josephson junctions formed from double-angle evaporated aluminum. The equivalent circuit schematic, shown in Fig.~\ref{fig:amp}b, displays a device with $C_\mathrm{a} = 1.085$ pF capacitance, $L_\mathrm{a} = 0.92$ nH linear inductance, and $L_\mathrm{J} = 0.516$ nH Josephson inductance.  When coupled to the 50-$\Omega$ input/output port with a capacitance of $C_\mathrm{c}=84$ fF, the amplifier resonance frequency (without flux bias) is at $4.028$ GHz with a quality factor of $20\times 10^{3}$. The Josephson inductance is realized as an array of $4$ superconducting quantum interference devices (SQUIDs). The SQUID loop areas are $5.5 \times 5 \ \mu$m$^2$, and the critical current of each SQUID is $3.2\  \mu$A. A high bandwidth microwave input line is coupled to the SQUID array with a mutual inductance of $80$ pH, allowing the microwave drive to modulate $L_\mathrm{J}$ at twice the amplifier frequency.  Additionally, residual capacitive coupling between the input line and the amplifier allows us to inject a small signal via this port. 

The device is wire-bonded to a microwave circuit board and cooled to 20~mK and protected from magnetic flux and electromagnetic radiation with associated shielding. The pump port of the device is filtered and attenuated (total of 50 dB attenuation). The device output passes through two cyrogenic circulators and amplified by a HEMT amplifier. We use an external coil to apply a dc flux bias to the system of approximately $\Phi_0/6$, where $\Phi_0$ is the magnetic flux quantum. As is shown in Fig.~\ref{fig:amp}c, this flux bias results in a linear coupling between the pump and the amplifier, allowing the pump to modulate the Josephson inductance at the pump frequency.  

We first characterize the amplifier response in the frequency domain. Figure~\ref{fig:amp}d displays the transmission through the amplifier when the pump tone is off (no gain) and when it is set to achieve a gain of approximately $4.2$ dB. The transmission exhibits a resonant response, with gain evident as an increase in the transmission. 

\begin{figure}
\centering
\includegraphics[width=0.5\textwidth]{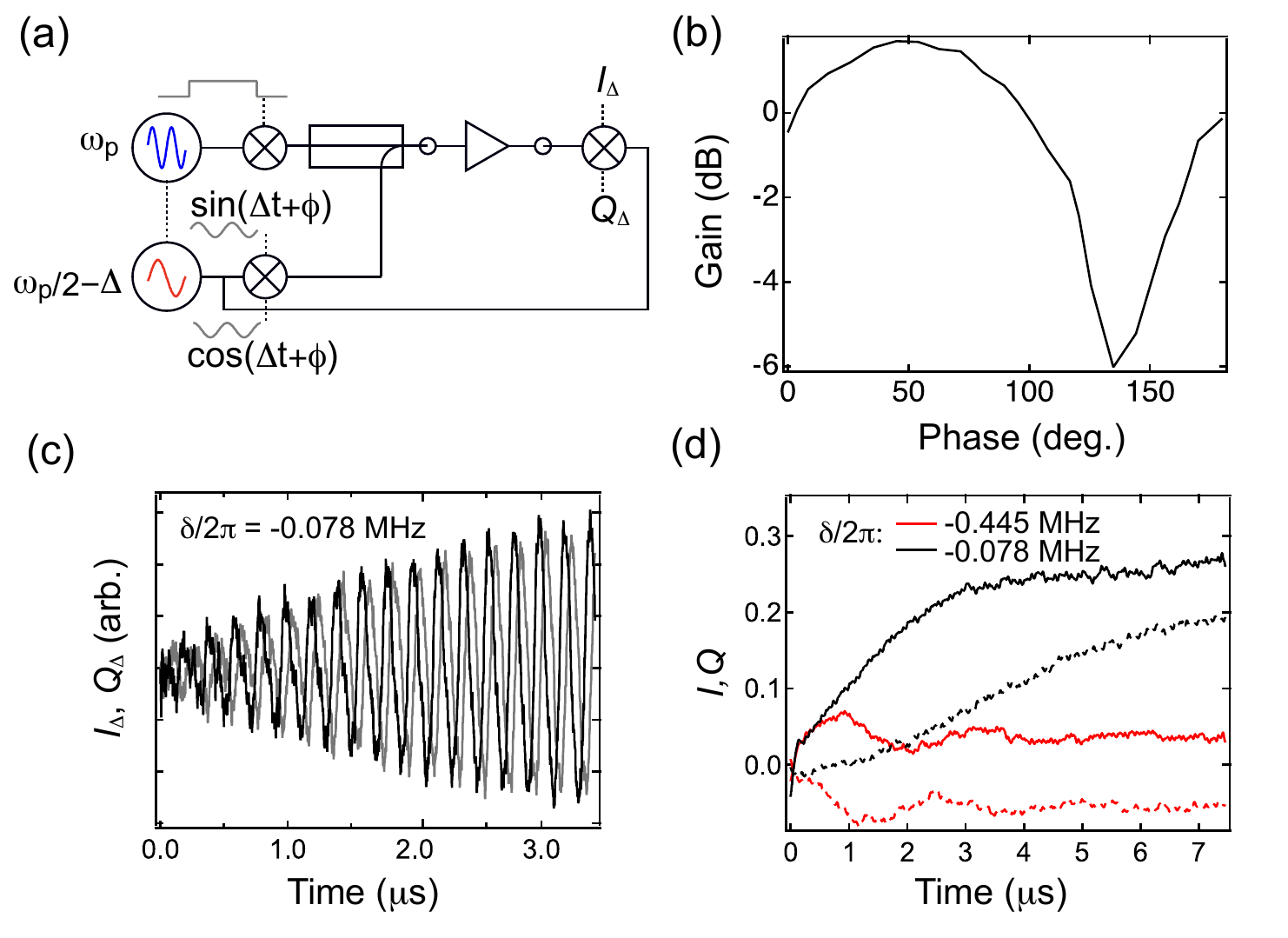}
\caption{{\bf Time domain response.} (a) The microwave setup includes a signal probe that is phase-locked to the pump frequency (dashed line). Single sideband modulation shifts the probe signal to be degenerate with the pump $\omega_\mathrm{s} = \omega_\mathrm{p}/2$. The output signal is demodulated. (b) The gain of the amplifier is phase-sensitive, we sweep the phase of the probe signal so that one quadrature is amplified and the other is squeezed. (c) Time domain response of the of $I_\Delta$ and $Q_\Delta$ where the signal and pump are turned on simultaneously. $I_\Delta$ and $Q_\Delta$ are demodulated to obtain $I$ and $Q$. (d) The time evolution of $I$ (solid)  and $Q$ (dashed) for two different values of the detuning.   }
\label{fig:timedomain}
\end{figure}

At this point, we have introduced a narrow-bandwidth parametric amplifier that is optimized for examining the interplay of gain/loss and coupling characteristic of the PT dimer.   To probe the PT-symmetry-breaking transition, we now turn to the time domain response of the amplifier. Figure \ref{fig:timedomain}(a) displays the microwave modulation/demodulation and timing of the experiment.  We utilize two (pump, signal) microwave generators operating at $\omega_\mathrm{p}$ and $\omega_\mathrm{s}-\Delta = \omega_\mathrm{p}/2-\Delta$, which are interferometrically locked via frequency doubling and demodulation to maintain exquisite phase stability. The signal generator's tone is upconverted by $\Delta$ using single-sideband modulation and combined with the pump tone. Signals that are transmitted through the amplifier are demodulated at frequency $\omega_\mathrm{s}-\Delta$, leading to heterodyne detection.

Since the amplifier is operated in degenerate mode ($2\omega_\mathrm{s} =\omega_\mathrm{p}$), the amplification is phase-sensitive, i.e., signals that are in-phase with the pump are amplified, and signals that are in quadrature with the pump are deamplified.  Figure \ref{fig:timedomain}(b) displays the amplifier transmission as the relative phase between the pump and signal is tuned, showing the characteristic response of phase-sensitive amplification. We choose the phase offset $\phi_0=45^\circ$ which corresponds to amplification of the input signal.

Figure \ref{fig:timedomain}(c) displays a representative time domain response of the amplifier. At $t=0$ the probe signal and pump are turned on, and the demodulated and digitized quadratures show an increasing envelope with modulation at the heterodyne frequency $\Delta/2\pi = 5$ MHz.  We further demodulate this heterodyne signal to determine the quadratures $I$ and $Q$. We use a Savitzky–Golay filter with a window size of 500 ns and polynomial order of 5 to smooth the data.  In Fig.~\ref{fig:timedomain}(d), we display the time response of $I,Q$ for different choices of the detuning between the signal/pump and amplifier resonance. The evolution of $(I,Q)$ exhibits a transition from oscillatory to saturated exponential dynamics.  This is the basis of the PT-symmetry-breaking transition which we now explore in further detail.

\begin{figure*}
\centering
\includegraphics[width=1\textwidth]{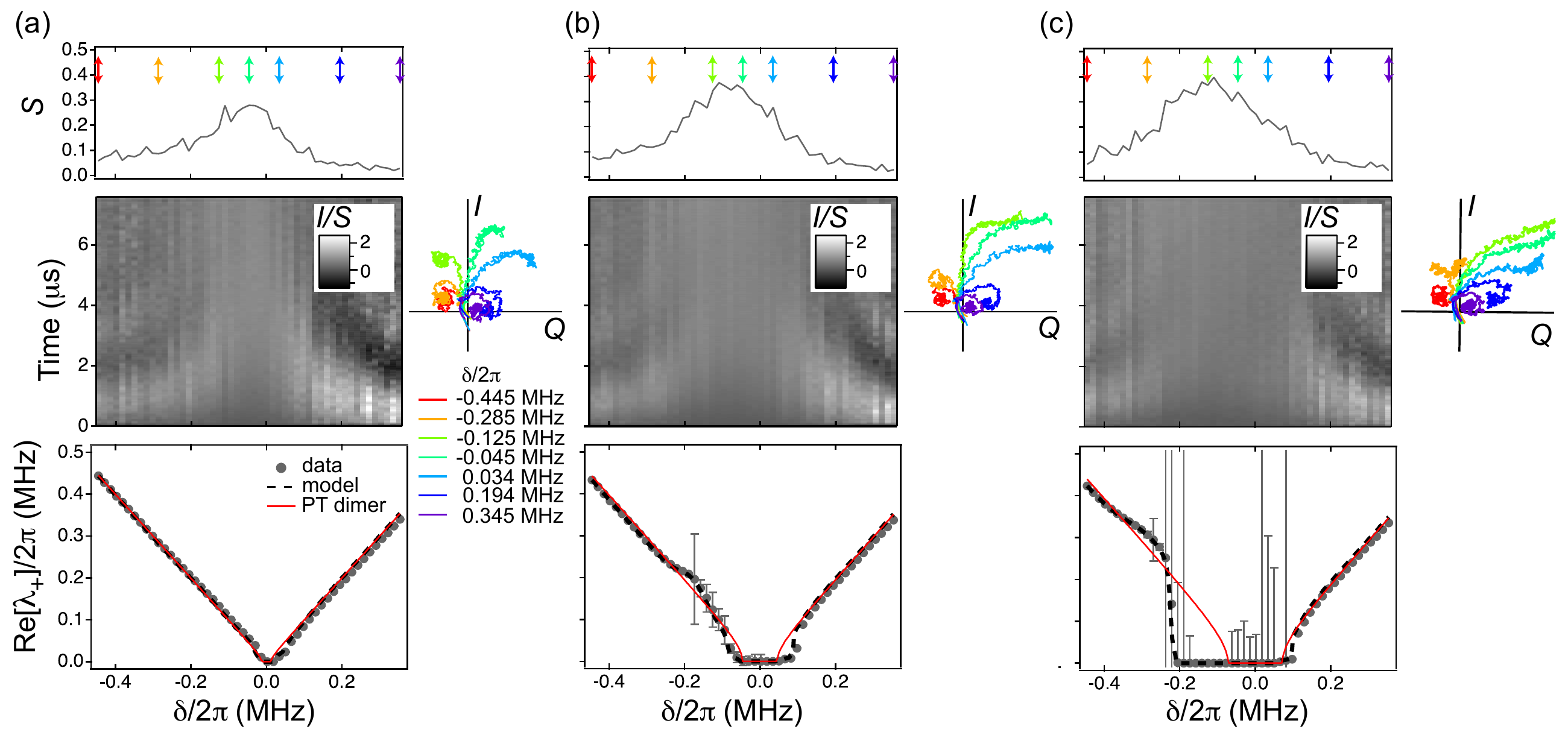}
\caption{{\bf PT-symmetry-breaking transition in a parametric amplifier.}  The graphs display the time response of $I$ versus $\delta$ for three different pump strengths, increasing sequentially corresponding to $\nu /2 \pi = 0.013$ (a), $0.046$ (b), and $0.072$ (c) MHz.  
The grayscale plots are normalized to the  steady state response (given by the value at a time duration of 7.6 $\mu$s, shown in the top panel). The side panels display the time evolution of $(I,Q)$ for a time duration of 7.512 $\mu$s. The lower panels display the extracted frequency from the transient response (markers). The error bars represent estimated errors of the fits and the dashed line gives the fit of the model's data. The red curves indicate the eigenvalues of the PT-dimer Hamiltonian (Eq.~\ref{eq:iqeom}).  A clear transition from oscillatory to amplified behavior corresponds to the PT-symmetry-breaking transition.   }
\label{fig:ptbreaking}
\end{figure*}

Figure \ref{fig:ptbreaking} displays the time response of $I$ versus detuning for three different pump strengths. For clarity, the grayscale plots are normalized to emphasize small amplitude features. The normalization factor ($S$) is depicted in the top panels.  Fig.~\ref{fig:ptbreaking}(a) displays the response at low pump amplitude, where the gain is small. We observe oscillation of $I$ in time, with a frequency consistent with the detuning.  As the pump strength is increased [Fig.~\ref{fig:ptbreaking}(b,c)] we observe a clear transition from oscillatory to non-oscillatory dynamics as the absolute value of the detuning is decreased. The evolution of $I$ vs. $Q$ for select detunings is shown in the associated side panels; these plots also highlight how there are values of the detuning where the dynamics are alternately oscillatory or non-oscillatory. 

  The observed data are in reasonable agreement with a simple model that takes into account the self-Kerr nonlinearity ($\chi$) and the probe signal at the Hamiltonian level,
\begin{equation}
    H = \delta a^{\dagger}a + i \frac{\nu}{2}(a^{\dagger 2}-a^{2})- \chi a^{\dagger 2}a^{2} +i \lambda(a^{\dagger}-a). \label{eq:fullmodel}
\end{equation}
Here, $\lambda$ is the probe signal strength. To model the amplifier dynamics, we initialize the oscillator in its ground state (i.e. $\ket{\psi (0)} = \ket{0}$) and let the state evolve under the influence of the Hamiltonian (Eq.~\ref{eq:fullmodel}). We model the effect of the output port by adding a Lindblad dissipator term $\sqrt{\kappa} a$, where $\kappa/2\pi =0.19$~MHz is determined from the measured the quality factor of the device. We use the Qutip \cite{Johansson2012,Johansson2013} master equation solver to obtain the time evolution of $I$ and $Q$ for different detunings. The input drive starts acting on the amplifier at $t=0$ at the same time three-wave mixing is also turned on. From the time evolution of the $I$ quadrature we fit the evolution to the function $A(t) = A_0 e^{-\alpha t}\sin(\omega t +\phi) + A_1$ to determine the frequency of oscillation $\omega$, which is displayed in the lowest panel. We apply the same analysis to the experimental data, where we seed the fit with initial guesses based on the results of the model. In practice, the fits which involve values of $\alpha$ and $\omega$ on the same order, are rather unconstrained (as indicated by the large estimated errors of the fits) and the analysis only demonstrates that the data is consistent with the model. We tune the model's  relative value of $\chi$ to achieve the best agreement with the experimental data. This fit yields a self-Kerr value of approximately $\chi / 2\pi =  0.095$ MHz and three different pump strengths as $\nu /2 \pi = 0.013$, $0.046$, and $0.072$ MHz. For comparison, we also display the associated eigenvalues of Eq.~\ref{eq:iqeom} based on the extracted values of $\nu$ from the model.   

The data exhibits asymmetry about zero detuning that arises due to the presence of the self-Kerr term in the Hamiltonian \eqref{eq:fullmodel}. The self-Kerr energy always decreases the resonance frequency, causing this term to dominate at the negative detuning values. Future work that strives to probe quantum correlations and entanglement in the output modes of the amplifier \cite{Wang2019} may have to adopt strategies to reduce the self-Kerr of such amplifier devices \cite{Fratini2018}.  

%

Our work establishes a connection between the transient dynamics of Josephson parametric amplifiers and the celebrated physics of the PT-symmetry-breaking transition.  In the context of optical devices, PT-symmetry and exceptional points have had a profound impact, demonstrating a broad range of devices and functionalities.  Future exploration of exceptional points with Josephson circuits can focus on quantum correlations and noise features in proximity to exceptional points, with applications in sensing and quantum information processing. 

\emph{Acknowledgements}---We thank R\'eouven Assouly, Yuxin Wang, Aashish Clerk, and Yogesh Joglekar for their helpful comments.  This research was supported by NSF Grant No.~PHY-1752844 (CAREER), the Air Force Office of Scientific Research (AFOSR) Multidisciplinary University Research Initiative (MURI) Award on Programmable systems with non-Hermitian quantum dynamics (Grant No.~FA9550-21-1-0202), ONR Grant No.~N00014-21-1-2630, and the Institute of Materials Science and Engineering at Washington University.


\end{document}